\documentclass{article}
\usepackage[final]{neurips_2021}

\usepackage{cite}
\usepackage{amsthm}

\usepackage{times}
\usepackage{bm}
\usepackage{natbib}
\usepackage{cite}
\usepackage{xcolor}
\usepackage{amsmath, amssymb}
\usepackage{graphicx}
\usepackage{url}
\usepackage{multirow}
\usepackage{cleveref}
\graphicspath{{mfl/figs}} 

\DeclareMathOperator*{\argmax}{arg\,max}

\def\d{\mathrm{d}}

\def\PP{\mathrm{pr}}

\def\tBP{\mathrm{3BP}}
\def\BeP{\mathrm{BeP}}
\def\hBP{\mathrm{h3BP}}
\def\MultiP{\mathrm{MultiP}}

\def\Pois{\mathrm{Poisson}}
\def\Binom{\mathrm{Binomial}}
\def\Bern{\mathrm{Bernoulli}}

\def\Betadist{\mathrm{Beta}}

\def\Multi{\mathrm{Multinomial}}

\def\ind{\bm{1}}

\newcommand{\aloc}{\psi} 
\newcommand{\acount}{x} 
\newcommand{\mcount}{X} 
\newcommand{\afreq}{\theta} 
\newcommand{\mfreq}{\Theta} 
\newcommand{\mass}{\alpha} 
\newcommand{\discount}{\sigma} 
\newcommand{\conc}{c} 
\newcommand{\ratemeas}{\nu} 

\newtheorem{theorem}{Theorem}

\newtheorem{remark}[theorem]{Remark}
\newtheorem{proposition}[theorem]{Proposition}
\newtheorem{assumption}{Assumption}

\usepackage{dsfont}
\usepackage{mathrsfs}

\def\E{\mathds{E}}

\def\de{\mathrm{d}}

\definecolor{forestgreen}{rgb}{0.0, 0.27, 0.13}

\title{Bayesian nonparametric strategies for power maximization in rare variants association studies}

\author{
  Lorenzo Masoero \\
  EECS, MIT\\
  \texttt{lom@mit.edu} 
    \And
  Joshua Schraiber\\
  Genome Interpretation Group, Illumina Inc\\
  \texttt{jschraiber@illumina.com} 
    \And
    Tamara Broderick \\ EECS, MIT \\ \texttt{tamarab@mit.edu}}

\begin{document}

\maketitle

\section{Introduction} \label{pbd:sec:intro}

Next generation sequencing, with larger genomic libraries and higher quality samples, has enabled scientists to start uncovering the genetic basis of  disease \citep{jackson2018genetic}. Studies of common genomic variants, like genome wide association studies (GWAS), have successfully helped researchers to identify the role of certain \emph{common} variants underlying disease  \citep{visscher2012five, visscher201710}. While GWAS have had an enormous impact on the understanding of the role of genetic basis in disease, it is now well accepted that common variants --- those targeted by GWAS --- only partially explain the biology and heritability of disease \citep{auer2015rare}. The large fraction of heritability unexplained by common variants is hypothesized to be largely caused by rare genomic variants \citep{pritchard2001rare, zuk2012mystery}. Hence, rare variants hold great promise: accurate study of their function could largely improve the  understanding of the biological underpinning of disease. 

The promise and value of studies focusing on rare variants is however hindered by the intrinsic challenges of the task. Rare variants are by their very definition present in few individuals: in order for association studies to reveal many rare variants, they must have large sample sizes, which might be costly. Moreover, extremely rare variants, such as singletons (genetic variants appearing in only  one individual within the study) are potentially the most interesting ones from a genetic perspective, but also the hardest to discover. In order to uncover these variants, expensive high quality samples --- collected with deep sequencing --- are required. Because sequencing studies are always conducted in the presence of limited budgets, devising procedures for optimally allocating this limited budget by tuning the sequencing parameters is an important, although challenging, open problem for practitioners. Indeed, the problem of optimal design in rare variants association studies [RVAS] has a rich history in the genomics literature
(\citet{ionita2011study, momozawa2020unique, cirulli2020genome}).  Recently, these types of considerations have been of interest also in the single-cell RNA sequencing community \citep{assefa2020utility, zhang2020determining}.

\textbf{Existing literature and our contribution:} Within the fast growing literature on rare variants, \citet{rashkin2017optimal} first considered the problem of optimal design of RVAS for power maximization of an associated statistical burden test. This important contribution provides practitioners with a quantitative framework for trade-offs evaluations when making complex design choices. However, the approach of \citet{rashkin2017optimal} also suffers from two main limitations: (a) it only applies to burden tests, and (b) does not directly apply to multi-stage optimal experimental design. In the present work, we provide a novel approach for optimal experimental design in RVAS, which tries to overcome the aforementioned limitations (a), (b). We develop a novel, rigorous statistical framework to understand how \emph{design} choices in RVAS can impact their usefulness, and we provide a practical workflow to inform such design choices under a fixed experimental budget. Our approach relies on the formulation of a rigorous statistical Bayesian nonparametric model. In particular, (a) our framework can help practitioners plan a power analysis in the context of any statistical test of interest, including but not limited to burden tests for rare variants, and (b) the Bayesian approach makes our framework automatically amenable to multi-stage optimal experimental design.
\section{Power trade-offs in RVAS via Bayesian nonparametrics} \label{pbd:sec:rvas}
 
\textbf{Rare variants association studies:}  Association studies test whether alleles are \emph{associated} with disease. This is done by performing a statistical test to assess if the frequency of a set of one or more alleles differs between healthy ``control'' subjects and  affected ``cases'' in a population of interest. Here we focus on RVAS, association studies designed to test whether \emph{rare} variants are associated with disease. One of the main challenges in RVAS, is that rare variants are hard to find and analyze: when \emph{designing} a RVAS, many factors should be carefully chosen in order to maximize the efficacy of the study --- the choice of the type of variants to analyze (e.g., disruptive vs.\ missense), the threshold for ``rare'' variants (e.g.\ observed in less than 1\% or 0.1\% of the samples), as well as the choice of the sequencing depth and samples sizes. Because each of these choices could impact the effectiveness of the study, practitioners must take care when designing the data collection process. In this work, we focus on the downstream impact of sequencing depth and sample sizes in RVAS. 

\textbf{Simple burden tests:} As in \citet{zuk2014searching}, we focus on power analysis for  a burden test for a binary trait (presence or absence of disease) in a two-class model, in which alleles are either null (i.e.\ abolishing function), or neutral (have no effect). We imagine having data from two groups, the affected ($A$) and unaffected ($U$) subpopulation, and assume that there exists a given, fixed and known, reference genome (an idealized complete genomic sequence), shared between the two subpopulations. We want to test whether \emph{rare} variants --- loci at which only a few individuals show deviation from the reference --- are correlated with the presence of disease. We do so by designing a statistical test that counts the abundance of rare variants in samples of affected and unaffected individuals, respectively. Given the reference and a user-specified threshold $\ell \in \{1,2,\ldots\}$, we let those variants appearing at most $\ell$ times in the sample be ``rare.'' Take subpopulations $A$ and $U$ with $N_A$ and $N_U$ samples, respectively. We define $\mu_{i}:=\mu_i(N_i,\ell)$ to be the (unknown) average value per individual of rare variants appearing in at most $\ell$ copies from $N_i$ samples from population $i \in \{A,U\}$. We test 
\begin{align}
    H_0  :  \{\mu_A = \mu_U\} \quad \text{vs.} \quad H_1  :  \{\mu_A > \mu_U\}. \label{pbd:eq:simple_burden_test}
\end{align}
\textbf{Test statistic and power:} To test $H_0$, we need a test statistic $T$, a function of the data testing $H_0$:
\begin{align}
    T = \frac{\bar{\mu}_A - \bar{\mu}_U}{\sqrt{s_A^2/M_A + s_U^2/M_U}}, \label{pbd:eq:two_sample_t_test}
\end{align}
where $M_i$ is the sample size of population $i$, $\bar{\mu}_i$ is the average number of $\ell$-tons in the sample per person in population $i$, $s_A^2, s_U^2$ are the sample variances of $\bar{\mu}_i$, $i \in \{A,U\}$. $T$ in \Cref{pbd:eq:two_sample_t_test} is the test statistic for a simple burden test; it counts the scaled excess of variants in $A$ relative to $U$. The \emph{power} of the test at a given confidence $\alpha \in (0,1)$ is the probability that $H_0$ is rejected at confidence $\alpha$ given that $H_0$ is false, $\pi_\alpha := \PP(H_0 \text{ rejected at confidence level } \alpha \mid H_0 \text{ false})$.

\textbf{Power maximization: optimal trade-off of sequencing depth and breadth under a fixed budget} We would like to design studies that help us detect association. Here, we adopt the power $\pi_{\alpha}$ of the associated burden test as a measure of detection efficacy. We investigate, in the presence of a fixed \emph{sequencing capacity} (budget), how sequencing depth and power of a simple burden test relate to each other: higher  depth comes with the promise of higher quality samples, in which genomic variants are called with higher precision, and in principle higher power. However, a higher sequencing depth also comes at higher cost. For a fixed budget, practitioners need to trade off the sequencing depth with the size of the cohorts sampled. 
Following \citet{rashkin2017optimal}, we let the cost of sequencing be $c(m, \lambda ; \kappa_0, \kappa_1) = m\lambda \kappa_1+ \kappa_0$. Here $m$ denotes the number of samples collected and $\lambda > 0$ is the average sequencing depth. The parameter  $\kappa_0$ is a library fixed cost, and $\kappa_1$ is a per-sample preparation cost. In the simplest case, which we consider in our experiments, $\kappa_0 = 0, \kappa_1  = 1$, and the cost of $m$ samples at depth $\lambda$ is simply $m\lambda$. $\kappa_0, \kappa_1$ could further depend on the population --- e.g., to encode the fact that obtaining samples from patients carrying a rare genetic disease might be more expensive than generic controls. Under a fixed budget $B>0$, we  need to trade off the depth $\lambda$ and the number $M_A, M_U$ of samples collected. Denoting the power as a function of the sequencing depth, as well as of the sample sizes, $\pi_{\alpha}(M_A, M_U, \lambda)$, we are therefore interested in solving
\begin{align}
    \argmax_{M_A, M_U, \lambda} \left\{\pi_\alpha(M_A, M_U, \lambda)\right\} \quad \text{subject to} \quad c(M_A+M_U,\lambda; \kappa_0, \kappa_1) \le B. \label{pbd:eq:power_maximization}
\end{align}

\textbf{A Bayesian nonparametric framework for rare variants association tests:} To solve \Cref{pbd:eq:power_maximization}, we briefly introduce a Bayesian nonparametric [BNP] framework (more details in  \Cref{pbd:sec:bnp,pbd:sec:bnp_hierar}). We assume the underlying \emph{reference} genome to be known. Let $N_A, N_U$ be the number of genomes collected in populations $A, U$: these are sequences of individuals, which can either agree (no variant) or disagree (variant) with the reference. Let variation be observed at $L < +\infty$ loci among the $N_A+N_U$ genomes collected in population $A, U$. We let $\Omega$ be an arbitrary measurable space of variant labels, and we let  $\aloc_\ell \in \Omega$ be the label of the $\ell$-th variant in order of appearance. We let $\acount_{i, n,\ell}$ equal $1$ if the variant with label $\aloc_\ell$ is observed for the $n$-th organism in population $i \in \{A,U\}$; otherwise, let $\acount_{i, n,\ell}$ equal $0$. We represent data for the $n$-th organism as a measure that pairs each variant count with the corresponding label: $\mcount_{i, n} := \sum_{\ell=1}^{L} \acount_{i, n,\ell} \delta_{\aloc_\ell}$. 

 We now posit a Bayesian model for the data $\bm{X}_i=\{\mcount_{i,1},\ldots,\mcount_{i,N_i}\}$, $i \in \{A,U\}$. We specify a likelihood function $\PP( \bm{X}_A,\bm{X}_U \mid \mfreq)$ and endow the latent parameter $\mfreq$ with a prior $\PP(\mfreq)$. Following \citet{masoero2018posterior,masoero2021more}, we adopt the (hierarchical) three-parameter beta-Bernoulli model. We imagine countably many latent variant rate-location pairs, collected in a random measure $\Theta_0= \sum_{\ell =1}^{+\infty}  \afreq_{0,\ell}\delta_{\aloc_{\ell}}$. The latent rates $\{\afreq_\ell\}\subset [0,1]$ are modeled as drawn from a three-parameter beta process \citep{teh2009indian, broderick2012beta}, parametrized by mass $\alpha > 0$, discount $\sigma \in [0,1)$, concentration $c>-\sigma$, denoted as $\bm{\xi}_0 = (\alpha,c,\sigma)$. For subpopulation $i  \in \{A,U\}$, every variant $\aloc_{\ell}$ has a population-dependent parameter $\theta_{i,\ell}$ governing variants occurrence probabilities:
\begin{align} \label{pbd:eq:h_freq}
    \theta_{i,\ell} \mid \Theta_0 \sim \Betadist\left\{a_i \theta_{0,\ell}, b_i(1-\theta_{0,\ell})\right\},
\end{align}
where $a_i >0, b_i>0$ are fixed parameters. For every individual $n$ in every population $i$, $\PP(\mcount_{n,i,\ell} = 1) = \afreq_{i,\ell}$: the probability that variant $\aloc_{\ell}$ is present in an individual varies in different populations, but depends on the underlying shared parameter $\theta_{0,\ell}$. Last, because of sequencing error, the actual observed presence or absence of variant $\aloc_\ell$ is modeled as ${Z}_{i, n,\ell} = \ind(C_{i, n,\ell,\text{noerror}} \ge D)\mcount_{i, n,\ell}$,
where $D>0$ is a fixed threshold, and for a fixed $p_{err} \in (0,1)$ quantifying a technology-dependent sequencing error parameter, $C_{i, n,\ell,\text{noerror}} \sim \Pois(\lambda (1-p_{err}))$, i.i.d.\ across $i,n,\ell$.

We leverage the model above to inform the trade-off between sample sizes and depth for power in rare-variants burden tests: the expectation and variance of the number of $k$-tons (number of variants appearing exactly $k$ times) predicted by our model in population $i \in \{A,U\}$ can be computed through the predictive structure implied by the model. Because this prediction explicitly depends on the sequencing depth, as well as the sample size, our model then captures the trade-off discussed above. Specifically, assuming $M_A$ and $M_U$ samples are collected from the affected and unaffected subpopulation respectively and fixing a common sequencing depth $\lambda>0$, we can replace the test statistic \Cref{pbd:eq:two_sample_t_test} with a  model-based counterpart to inform the power trade offs. Namely:

\textbf{Q1}: Analyze the power of the rare variants burden test as we increase the sample size of the cases and controls, for a fixed sample size, for different choices of the sequencing depth $\lambda$.

\textbf{Q2} Analyze, under a fixed budget, how different sequencing depths and sample sizes affect the power of the burden test, by replacing the test statistic $T$ (Eq. \ref{pbd:eq:two_sample_t_test}) with its model-based counterpart.

\section{Experiments} \label{pbd:sec:experiments_hierar}

\begin{figure}
    \centering
    \includegraphics[width=\textwidth,height=\textheight,keepaspectratio]{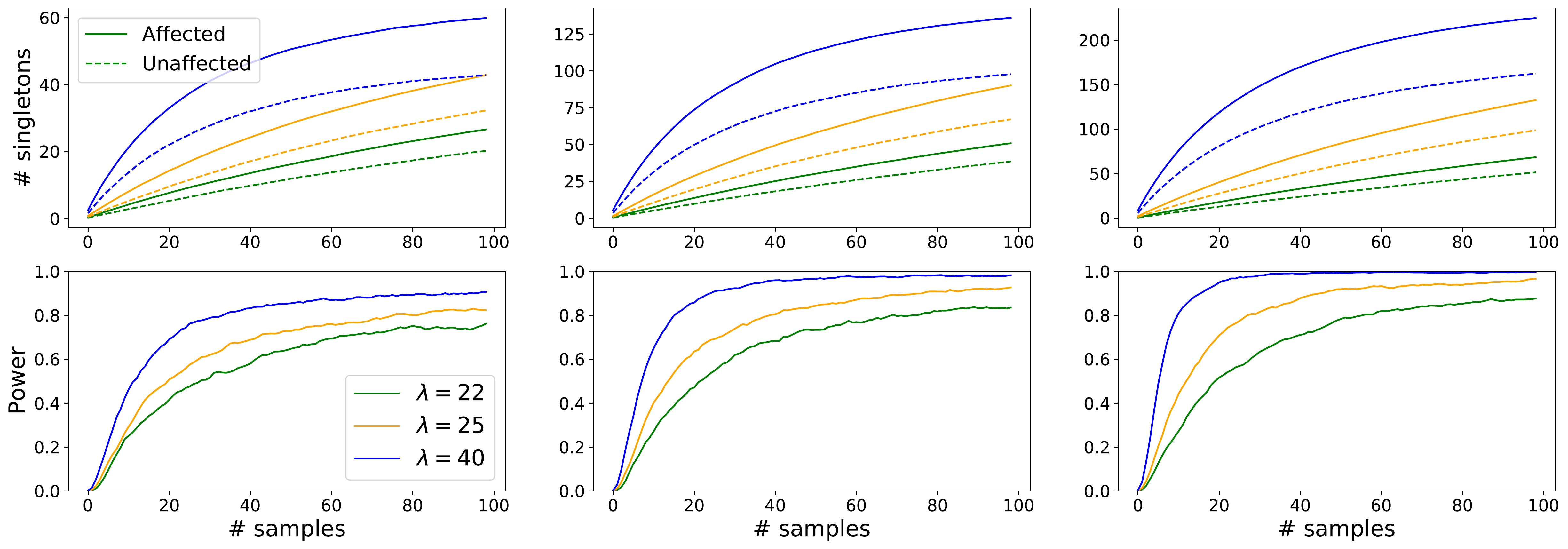}
    \caption{Top row: number of singletons in the affected (solid) and unaffected (dashed) subpopulations (vertical axis), for various sequencing depths (green: $\lambda = 22$, orange: $\lambda = 25$, blue: $\lambda = 40$, horizontal axis). Bottom row: power of the  test (vertical axis) for a given sample size (horizontal axis). Different columns refer to different  parameters of $\mfreq_0$ --- left: $\bm{\xi}_0 = [5,4,0.5]$, center: $\bm{\xi}_0 = [8,5,0.55]$, right: $\bm{\xi}_0 = [10,6,0.6]$. For all subplots, $a_1 = 200, a_2 = 150, b_1, b_2 = 100$.}
    \label{pbd:fig:hibp_fixed}
\end{figure}

We now move on to the empirical evaluation of the properties of the Bayesian hierarchical model via simulated data from the model. Additional details can be found in \Cref{pbd:sec:experiments,pbd:sec:experiments_hierar_2}. We  address Q1 and Q2 introduced at the end of \Cref{pbd:sec:rvas}. For Q1, in \Cref{pbd:fig:hibp_fixed} we show how the power of the singleton burden test changes with the sizes of the control and the affected subpopulations, for fixed sequencing depth. Larger sample sizes and higher sequencing depths never harm the power. However, the rate at which power increases depends on the underlying parameters of the model; in real data the same sequencing strategy could prove more or less effective depending on the data. If the experiment under study has the goal of achieving a desired level of power, our framework would allow practitioners to provide estimates of the budget needed in order to achieve the desired power. 

\begin{figure}
    \centering
    \includegraphics[width=\textwidth,height=\textheight,keepaspectratio]{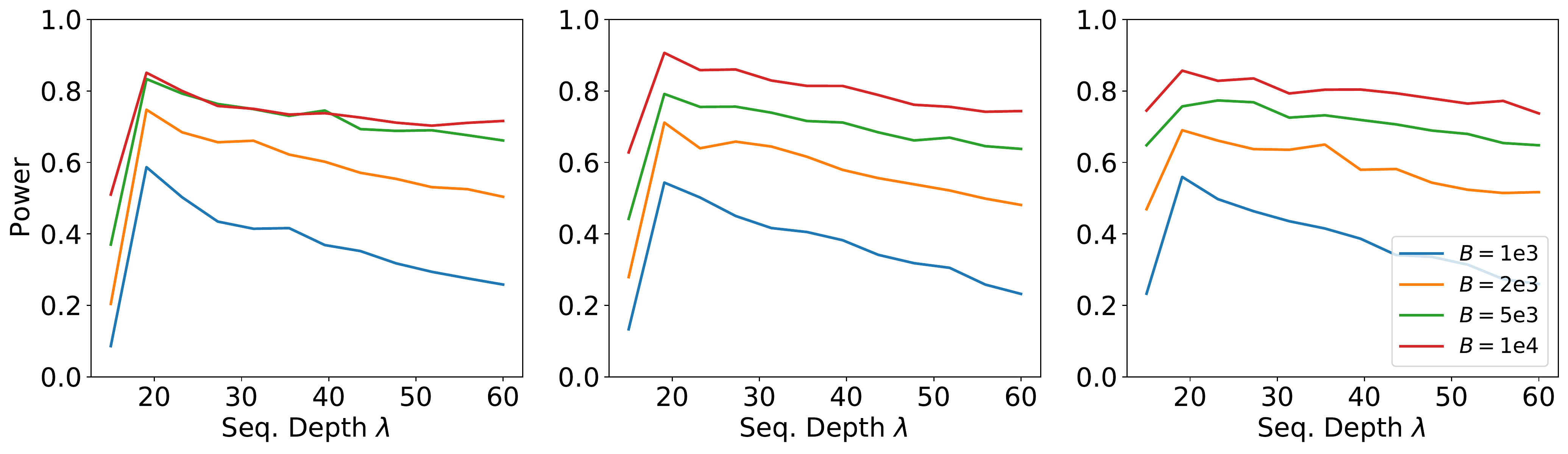}
    \caption{Power of the singletons hypothesis burden test for $H_0$ (vertical axis) as the sequencing depth changes (horizontal axis) under a fixed sequencing capacity (i.e.\ budget) constraint (Q2). Different lines report results for different budgets. Different columns refer to different choices of the hyperparameters of the underlying shared measure $\mfreq_0$ --- left $\bm{\xi}_0 = [5,4,0.5]$, center $\bm{\xi}_0 = [8,5,0.55]$, right $\bm{\xi}_0 = [10,6,0.6]$. For all subplots, $a_A = 200, a_U = 150, b_A = 100, b_U = 100$.}
    \label{pbd:fig:hibp_budget}
\end{figure}

For Q2, power maximization under a fixed budget, we consider singletons --- variants appearing at most once (see \Cref{pbd:sec:experiments_hierar_2} for $k$-tons, $k \in \{2,3,4,5\}$). In line with \citet[Figures 4,5]{rashkin2017optimal}, we find that the optimal sequencing depth is relatively insensitive to the available budget (\Cref{pbd:fig:hibp_budget}). Moreover, we find that tests for extremely rare variants generally achieve higher power than tests for relatively less rare variants (\Cref{pbd:fig:hibp_budget_ktons}). The intuition has to be sought in the properties of the model's frequency distribution: the underlying three-parameter beta process prior used here suggests that most variants are extremely rare. Therefore, when testing for extremely rare variants, the test has sufficient power even when the sequencing depth is relatively low, just because on average the affected population will reveal a larger number of singletons with respect to the unaffected population. When testing for relatively more-frequent variants, a larger depth is needed in order to capture a significant discrepancy between the affected and unaffected population.

\section{Discussion} \label{pbd:sec:discussion}

We introduced a novel BNP approach for optimal design for power maximization in RVAS. We showed with simulations that our model can be a useful tool to inform complex experimental design choices, especially under a fixed budget constraint. We envision several extensions to our work:

\textbf{Posterior analyses}: Our method could be  used for multi-stage sequencing design, i.e.\ when scientists have existing data, and have to optimally design future experiments \citep{pahl2009optimal}.

\textbf{Beyond simple burden tests}: Because of the statistical framework for the data generating process, our method can be used for other tests (e.g., Sequence Kernel Association Test \citep{wu2011rare}).

\textbf{Variant calling rules}: We here followed  \citet{ionita2010optimal} and used a simple threshold rule for variant calling. Evaluating sensitivity variant callers, and producing results for genotype variant calling rules \citep{nielsen2011genotype} could further enhance the usefulness of our approach.

\paragraph{Acknowledgments} Lorenzo Masoero
and Tamara Broderick were supported in part by the DARPA I2O LwLL
program, an NSF CAREER Award, and ONR.
\bibliographystyle{abbrvnat}
\bibliography{refs.bib}
\appendix
\section{Simple burden tests: additional details}  \label{pbd:sec:burden_tests}

In this work, following \citet{zuk2014searching}, we focus on the simplest possible power analysis for rare variants association studies: we consider burden tests for a binary trait (presence or absence of disease) in a two-class model, in which all alleles are either null (i.e.\ abolishing gene function), or neutral (have no effect on gene function). That is, we imagine having data from two groups, an affected ($A$) and unaffected ($U$) subpopulation, and design a (statistical) burden test with the goal of understanding if certain rare genomic variants are correlated with the presence of disease by counting their abundance in samples of affected and unaffected individuals respectively. In particular, we imagine that there exists a given, fixed and known, reference genome, shared between the two subpopulations. And we interested in testing for \emph{rare} variants. Here, given the reference and a user-specified threshold $\ell \in \{1,2,\ldots\}$, we let ``rare'' variants be those variants appearing at most $\ell$ times, with $\ell \ge 1$ in the sample. Take subpopulations $i$ and $j$, with $i\ne j$. Assume subpopulations $i$ and $j$ have $N_i$ and $N_j$ samples, respectively. Then we define
\begin{itemize}
    \item $\mu_{i}:=\mu_i(N_i,\ell)$ to be the (unknown) average value per individual of rare variants appearing in at most $\ell$ copies from $N_i$ samples from population $i$
    \item $\mu^\star_i :=\mu^\star_i(N_i,N_j,\ell)$ to be the (unknown) mean value per individual of rare variants appearing at most in $\ell$ copies from $N_i$ samples from population $i$, and \emph{not} appearing in any of $N_j$ samples from population $j$
\end{itemize} 
Variants are computed with respect to a known reference genome, shared by the two subpopulations. We then want to test the hypotheses
\begin{align}
    H_0  :  \{\mu_A = \mu_U\} \quad \text{vs.} \quad H_1  :  \{\mu_A > \mu_U\}, \nonumber
\intertext{and}
    H_0^\star :  \{\mu_A^\star = \mu^\star_U\} \quad \text{vs.} \quad H_1^\star  :  \{\mu^\star_A > \mu^\star_U\}.
\label{pbd:eq:simple_burden_test_unique}
\end{align}
A special instance of this hypothesis test --- considered in \citet{rashkin2017optimal} --- is the one in which $\ell = 1$, i.e., those variants that have only been observed once in the sample.  We henceforth investigate, in the presence of a fixed \emph{sequencing capacity}, how sequencing depth and power of a simple burden test relate to each other. And specifically,
\begin{itemize}
    \item Given a fixed experimental design, and potentially an existing study, how will the power of the burden test change as we increase the sample sizes of the affected and unaffected samples?
    \item Given a fixed budget, and potentially an existing study, how should practitioners design a follow-up study to maximize the power of a simple burden test, trading off depth of the sequencing and size of the affected and unaffected cohorts?
\end{itemize}

\subsection{Test statistic, null distribution, power}
We now introduce a statistical framework for power analysis in the simple burden test introduced in \Cref{pbd:eq:simple_burden_test}, building on the approach of \citet{rashkin2017optimal}.

We test the null hypothesis $H_0$ in \Cref{pbd:eq:simple_burden_test}. The same assumptions made for the null hypothesis \Cref{pbd:eq:simple_burden_test} are also made for the hypothesis test in \Cref{pbd:eq:simple_burden_test_unique} --- we here omit the statements. 

First, for testing $H_0$, we compute the test statistic already defined in \Cref{pbd:eq:two_sample_t_test}:
\begin{align}
    T = \frac{\bar{\mu}_A - \bar{\mu}_U}{\sqrt{s_A^2/M_A + s_U^2/M_U}},
\end{align}
where 
\begin{itemize}
    \item $M_A, M_U$ are the sample sizes of group $A, U$ respectively
    \item $\bar{\mu}_A, \bar{\mu}_U$ are the average number in the sample of $\ell$-tons per person in $A, U$ respectively
    \item $s_A^2, s_U^2$ are the sample variances of $\bar{\mu}_A, \bar{\mu}_U$ respectively
\end{itemize} 

\begin{assumption}[Distribution under the null] \label{pbd:assump:null}
    Under $H_0$, we assume that $\bar{\mu}_A, \bar{\mu}_U$ are normally distributed with the same mean, but potentially unequal variances. Then, the test statistic $T$ in \Cref{pbd:eq:two_sample_t_test} follows a (centered) Student's-$t$ distribution with (approximate) degrees of freedom given by the Welch-Satterwaite equation \citep{welch1947generalization},
    \[
        \phi_0 = \frac{\left(\frac{s_A^2}{M_A}+\frac{s_U^2}{M_U}\right)^2}{\frac{s_A^2}{M_A^2(M_A-1)}+ \frac{s_U^4}{M_U^2(M_U-1)}},
    \]
    i.e.\
    \[
        T \mid H_0 \sim \tau(\phi_0, 0).
    \]
\end{assumption}
Hence, for a given confidence level $\alpha \in [0,1]$, we reject the null hypothesis $H_0$ if the test statistic is sufficiently large, namely $ T>t_{1-\alpha}(\phi_0, 0)$ --- where $t_{1-\alpha}(\phi_0, 0)$ is the $(1-\alpha)\times 100$\% percentile of a (centered) $t$-distribution with $\phi_0$ degrees of freedom, i.e. $\PP(T > t_{1-\alpha}(\phi_0, 0) \mid H_0) = \alpha$.
\begin{assumption}[Power of two-sample $t$-test] \label{pbd:assumption:power}
    The power of the test,
    \begin{align}
        \pi_\alpha(M_A, M_U) := \PP(H_0 \text{ rejected at confidence level } \alpha \mid H_0 \text{ false})
    \end{align}is computed as
    \begin{align} \label{pbd:eq:power_t_test}
        \pi_\alpha(M_A, M_U) = \PP\left[X > t_{1-\alpha}(\phi_0, 0) \mid X \sim \tau(\phi_0, T)\right],
    \end{align}
    where $t_{\alpha/2}(\phi_0, 0)$ is the $(1-\alpha/2)\times 100\%$ percentile of the (central) $t$-distribution with $\phi_0$ degrees of freedom, while $\tau(\phi_0,\lambda)$ is the law of a random variable following a non-central Student's-$t$ distribution with $\phi_0$ degrees of freedom and noncentrality parameter $\lambda$ \citep{harrison2004sample}.
\end{assumption}
\section{A Bayesian nonparametric framework for power maximization} \label{pbd:sec:bnp}

In this section, we provide additional details and background on the Bayesian nonparametric approach used in \Cref{pbd:sec:rvas} to address the trade-off discussed in \Cref{pbd:sec:rvas}. In this section, we consider a simple nonparametric model, in which (i) the arising of variants in different subpopulations is modeled independently, and (ii) we treat genomes as haploid sequences. While simplistic, this model allows us to build intuition. Next, in \Cref{pbd:sec:bnp_hierar}, we introduce a hierarchical generalization of our model, so as to be able to model the arising of diploid variants in multiple subpopulations jointly. 

Our present model, and its justification, coincide with \citet{masoero2021more} --- there in the context of genomic variants prediction. We here briefly describe it, by considering the case of a single subpopulation in the simpler case in which the data is collected without any sequencing error. 

\subsection{Observational model without sequencing errors} \label{pbd:sec:model_no_noise}

With the advent of next-generation sequencing, it is now possible to determine with high precision an organism's genome \citep{reuter2015high}. When studying a population of interest, a first step towards the understanding of the landscape of variation within the population is the definition of a \emph{reference} genome: a fixed representative sequence for the whole population. While several forms of variants exist (missense, truncation, translocation, etc.), in the present work we treat all types of variation from the reference as equal. 

Let $N$ be the number of genomes collected: these are (genomic) sequences of individuals, which can either agree (no variant) or disagree (variant) with an underlying fixed reference genome at a large, finite number of loci. Let $L$ be the number of loci at which variation is observed among the first $N$ genomes collected, $0 \le L < +\infty$. We let $\Omega$ be an arbitrary measurable space of variants labels, and $\aloc_\ell \in \Omega$ be the label of the $\ell$-th variant in order of appearance, and $\acount_{n,\ell}$ equal $1$ if the variant with label $\aloc_\ell$ is observed for the $n$-th organism; otherwise, let $\acount_{n,\ell}$ equal $0$. We collect data for the $n$-th organism in  a measure which pairs each variant observation with the corresponding variant label: $\mcount_{n} := \sum_{\ell=1}^{L} \acount_{n,\ell} \delta_{\aloc_\ell}$. 

Given the data $\mcount_1,\ldots,\mcount_{N}$, we now posit a Bayesian model for it. This requires us to specify a generative model for the data, via a likelihood function $\PP(\mcount_{1}, \ldots, \mcount_{N}\, |\, \mfreq)$, and endow the latent parameter $\mfreq$ with an adequate prior distribution $\PP(\mfreq)$. Following \citet{masoero2021more}, we here adopt the three-parameter beta-Bernoulli model. In this model, we imagine an underlying countable infinity of latent variants rate-location pairs, labelled as $\{ ( \afreq_\ell,\aloc_\ell ) \}_{\ell\geq 1}$, which are the realization of a Poisson point process, whose distribution is characterized by the following rate measure on $[0,1]\times \Omega$:
\begin{align} \label{pbd:eq:rate_meas}
\ratemeas(\d \afreq)P_0(\de\omega)
 		 = \mass \frac{\Gamma(1+\conc)}{\Gamma(1-\discount)\Gamma(\conc+\discount)} \afreq^{-1-\discount} (1-\afreq)^{\conc+\discount-1} \bm{1}_{[0,1]}(\afreq)\d \afreq P_0(\de\omega),
\end{align}
where $\bm{1}_A(\afreq)$ is equal to $1$ if the argument $\afreq$ belongs to the set $A$, and $0$ otherwise. Here $P_0$ is an arbitrary diffuse measure on the space $\Omega$, and will be irrelevant in our discussion. We henceforth ignore it, and write $\mfreq \sim \tBP(\mass, \discount, \conc)$ to denote a Poisson point process with L\'{e}vy mean measure as above. If we impose $\mass>0, \discount \in [0,1), \conc > -\discount$, then $\mfreq$ can be seen as a discrete random measure on $\Omega$, with countably many atoms, and such that the sum of the rates is finite. That is, we can represent $\mfreq$ as follows:
\[
    \mfreq= \sum_{\ell\geq 1} \afreq_\ell \delta_{\aloc_{\ell}}.
\]
By construction, the rates $\{\afreq_\ell\}_{\ell \ge 1}$ are all values in $[0,1]$: we interpret then each rate $\afreq_\ell$ as the probability that the corresponding variant $\aloc_\ell$ is present in an individual, independently of any other individual and variant. I.e., conditionally on $\mfreq$, we let $X_n$ be a Bernoulli process with underlying rate measure $\mfreq$ --- $\mcount_n\sim\BeP(\mfreq)$. We write $\mcount_{n} := \sum_{\ell\geq1} \acount_{n,\ell} \delta_{\aloc_\ell}$; since $\acount_{n,\ell} = 0$ for all unobserved variants, this equation reduces to the previous definition of $\mcount_n$ above. Then, the model  allows the observed number of variants to be finite for any finite dataset, and grow without bound as the number of observations increases \citep{teh2009indian, james2017bayesian, broderick2018posteriors}. To sum up:
\begin{align}
    \mfreq \sim \tBP(\mass, \conc, \discount), \quad \mcount_n\mid\mfreq \sim \BeP(\mfreq), \quad n=1,\ldots,N. \label{pbd:eq:model_bnp}
\end{align}

\subsection{Observations in the presence of sequencing errors} \label{pbd:sec:model_noise}
The model given in \Cref{pbd:sec:model_no_noise} describes a data generating process that does not take into account the presence of sampling error in the sequencing procedure. I.e., it implicitly assumes that variants are detected without error whenever present. However, in practice, sequencing is a complex and noisy process, in which millions of reads of fragments of the same genomic sequence need to be aligned and compared to the reference genome. In this process, several mistakes can be made when reconstructing the genomic sequence.

Now, following again \citet{masoero2021more}, we introduce a modification of our model that takes into account the error induced by sequencing.
 Let $X_n$ be the underlying, true value of the binary genotype sequence associated with individual $n$. We assume that, in the presence of sequencing errors, observations are obtained by down-weighting the probability of actually observing a variant:
\begin{enumerate}
    \item For each individual $n$, each locus in the sequence is read a random number of times $C_{n,\ell}$. Given a sequencing depth $\lambda>0$, we assume
    \begin{align*}
     C_{n,\ell} \overset{\rm iid}{\sim} \Pois(\lambda).
    \end{align*}
    \item Each of these reads is either correctly processed, or incurs some form of reading error. We assume that 
    \begin{align*}
     C_{n,\ell,\text{noerror}} \sim \Binom(C_{n,\ell}, 1-p_{err}).
    \end{align*}
    Here $p_{err} \in [0,1)$ is a technology-dependent, known parameter. 
    \item We let $D$ be a threshold parameter for a simple variant calling rule --- fixed a priori. Then, a variant at locus $\ell$ for individual $n$ is called according to the following variant-calling rule:
    \begin{align}
        {Z}_{n,\ell}(\lambda, D, p_{err}) = \ind(C_{n,\ell,\text{noerror}} \ge D)X_{n,\ell}. \label{pbd:eq:variant_caller}
    \end{align}
\end{enumerate}

\begin{remark}
    Given steps 1.---3. above, the probability of obtaining at least $D$ correct reads from the Poisson-binomial sampling described is given by
        \begin{align*}
            \phi:=\phi(\lambda, D, p_{err}) &= \sum_{d \ge D} \frac{e^{-\lambda}\lambda^d}{d!} \sum_{i = D}^d \binom{d}{i}(1-p_{err})^i p_{err}^{d-i} \\
            &= \sum_{d\ge D} \frac{e^{-\lambda(1-p_{err})}\{\lambda(1-p_{err})\}^d}{d!}.
        \end{align*}
        Then, it follows that for all $n, \ell$, $Z_{n,\ell} \mid \afreq \sim \Bern(\phi\afreq_\ell)$. I.e., 
    \begin{align}
        Z_n(\lambda, D, p_{err}) \mid \afreq \sim \BeP(\phi\afreq). \label{pbd:eq:model_error}
    \end{align}
\end{remark}

As discussed in \citet{masoero2021more}, the simple threshold variant calling rule used here is a simplification of modern variant callers employed in genomic pipelines (see \citet{xu2018review} for a review of variant calling algorithms). Exploring how to embed alternative variant calling rules within the present formulation is an exciting direction for future work, as discussed in \Cref{pbd:sec:discussion}.

\subsection{Distributional results from the model} \label{pbd:sec_app:bnp_uni}
After introducing a model for sequencing in the presence of noise in \Cref{pbd:sec:model_noise}, we now complete our derivation so as to be able to use the model for the problem of optimal experimental design of burden tests under a fixed budget, previously discussed in \Cref{pbd:sec:intro}. Specifically, we characterize the predictive behavior for the number of rare variants induced by the model. This quantity can be used in order to solve the power maximization problem discussed in \Cref{pbd:eq:power_maximization}. In particular, we first predict how  many new rare variants are going to be discovered in future samples, as a function of the sequencing depth and the extrapolation size. As a direct consequence, we can use these predictions to analyze how the power of the corresponding burden test is going to be affected, and in turn aid experimental design. We now state the key distributional results implied by the model. See \citet[Proposition A.3]{masoero2021more} for a proof of this proposition.

 \begin{proposition} \label{pbd:prop:k_tons_error} 
    Let $Z_{1},\ldots, Z_{N}$ be $N \ge 0$ training draws from the Bayesian nonparametric model \Cref{pbd:eq:model_error}, collected at sequencing depth $\lambda_{pilot}>0$, for a fixed variant calling threshold $D \in \mathbb{N}$ and sequencing error $p_{err} \in [0,1]$, 
    \[
        Z_n \mid \afreq, \lambda_{pilot}, D, p_{err} \overset{\rm iid}{\sim} \BeP(\phi\afreq), n = 1,\ldots,N \quad \text{and } \afreq \sim \tBP(\mass, c, \discount).
    \]
    For a given frequency $k \in \{1,\ldots,M\}$, the number $U_N^{(M,k)}$ of variants appearing exactly $k$ times in $M$ additional samples now at adjusted sequencing depth $\lambda_{follow}>0$, conditionally on the first $N$ observations $Z_{1:N}$ is given by
    \[
        U_N^{(M,k)} \mid Z_{1:N}, \lambda_{pilot}, 
        \lambda_{follow}, D, p_{err}  \sim \Pois\left( \gamma_k \right),
    \]
    where, letting $\phi_{pilot} = \phi(\lambda_{pilot}, D, p_{err})$ and $\phi_{follow} = \phi(\lambda_{follow}, D, p_{err})$
    \begin{align}
     \gamma_k&:=\gamma_k(N, M, \phi_{pilot}, \phi_{follow}, \bm{\xi}) \label{pbd:eq:bnp_newz}\\
        &= \mass \binom{M}{k} (\phi_{follow})^k \frac{(1+c)_{k-1\uparrow}}{(1-\discount)_{k-1\uparrow}} \E\left[ (1-\phi_{follow}B)^{M-k}(1-\phi_{pilot}B)^N\right]. \nonumber
    \end{align}
    Here, $B \sim \Betadist(k-\discount,c+\discount)$, and $(a)_{b \uparrow}:=\Gamma(a+b)/\Gamma(a)$. Under our model, observations are exchangeable --- equivalently, conditionally i.i.d.\ given the underlying random parameter $\afreq$. Then, conditionally on past samples $Z_{1},\ldots,Z_N$, the number of new variants per sample which appear exactly $k$ times in the $M$ new samples must be distributed as
    \begin{align}
        S_N^{(M,k)} \mid Z_{1:N} \sim \Pois\left( \frac{\gamma_k}{M}\right). \label{pbd:eq:news_single_error}
    \end{align}
 \end{proposition} 

\subsection{Power analysis: prior analyses and follow-up designs} \label{pbd:sec:depth_power_bnp}
We have now all the ingredients to illustrate how the Bayesian formulation proposed above can aid optimal experimental design, when the goal is to maximize the power of a burden test for rare variants like the one presented in \Cref{pbd:eq:two_sample_t_test}. In what follows, we will focus on the burden test introduced in \Cref{pbd:sec:rvas}.

We consider again the case of two subpopulations --- the affected (``$A$'') and unaffected (``$U$'') respectively. Given hyperparameters $\bm{\xi}_j = [\mass^{(j)}, \conc^{(j)}, \discount^{(j)}]$ for $j\in\{A,U\}$, we  leverage the predictive behavior given in  \Cref{pbd:eq:news_single_error} to inform the trade-off between sample sizes and depth for power in rare-variants burden tests. In particular, the expected number of $k$-tons (number of variants appearing exactly $k$ times) predicted by our model in population $j \in \{A,U\}$, is directly computed from \Cref{pbd:prop:k_tons_error} for the special case $N=0$. And since this prediction explicitly depends on the sequencing depth, as well as the (total) sample size, our model then captures the trade-off discussed in \Cref{pbd:sec:intro}. Given hyperparameters $\bm{\xi}_j$, let $\gamma_k^{(j)}:=\gamma_k(0,M,1, \phi_{j}, \bm{\xi}_j)$ be the expected number of $k$-tons in $M$ samples in population $j$ (see \Cref{pbd:eq:bnp_newz}). Then, assuming $M_A$ and $M_U$ samples are collected from the affected and unaffected subpopulation respectively, and fixing a common sequencing depth $\lambda>0$, we compute 
\begin{align}
    T(M_A, M_U, \lambda \mid \bm{\xi}_A, \bm{\xi}_U) = \frac{\frac{\gamma_k^{(A)}}{M_A} - \frac{\gamma_k^{(U)}}{M_U}}{\sqrt{\frac{\gamma_k^{(A)}}{M_A^2} + \frac{\gamma_k^{(U)}}{M_U^2}}}. \label{pbd:eq:model_based_test}
\end{align}
$T$ above is a model-based counterpart of the burden test statistic given in \Cref{pbd:eq:two_sample_t_test}. Then, under the assumptions in \Cref{pbd:sec:burden_tests}, and using the additional approximation that for $\mu$ large, the distribution of $X \sim \Pois(\mu)$ is well approximated by a Gaussian distribution with mean and variance $\mu$,  we can use \Cref{pbd:eq:model_based_test} to provide a model-based answer to both questions Q1 and Q2 in \Cref{pbd:sec:intro}. Namely, we can: \begin{itemize}
    \item[Q1:] Analyze how the power of the rare variants burden test changes as we increase the sample size of the cases and controls, for a fixed sample size. Under our model, this analysis will depend on (i) the underlying hyperparameters $\bm{\xi}_j$ governing the rare variants distributions, and (ii) the choice of the sequencing depth $\lambda$.
    \item[Q2:] Analyze, under a fixed budget, how different choices of \emph{feasible} sequencing depths and sample sizes will affect the power of the burden test. In particular, we do so by leveraging \Cref{pbd:assumption:power}, where we now replace the test statistic $T$ in \Cref{pbd:eq:power_t_test} with its model-based counterpart defined in \Cref{pbd:eq:model_based_test}.
\end{itemize}
While in the present manuscript we focus on prior analyses, we emphasize that the model could also be helpful for the design of a follow-up study, when the goal is to maximize the usefulness of budget allocation in a multi-stage experiment. That is, given pilot data from both the affected and unaffected subpopulations, we could first infer the underlying parameters $\bm{\xi}_A, \bm{\xi}_U$ from the data (e.g., via an empirical Bayes procedure), and then, using these inferred values, design a follow-up study with the goal of maximizing the power of a burden test that is using future samples.

\section{Experiments} \label{pbd:sec:experiments}

We now present experimental results using the model described in \Cref{pbd:sec:bnp}. In our experiments, we perform \emph{prior} analyses, and report results similar to the ones obtained by \citet{rashkin2017optimal}. Specifically, we first perform a thorough investigation of how different configurations of the model's hyperparameters affect the data generating process in \Cref{pbd:sec:exp_viz}. Then, we move on to the power analysis trade-offs in \Cref{pbd:sec:exp_pow}. Throughout our experiments, we fix the threshold for the variant calling rule to $D = 30$, and the error probability to $p_{err} = 0.05$. 
 
\subsection{New, rare and excess variants in the Bayesian nonparametric model} \label{pbd:sec:exp_viz}

We here provide a thorough analysis of synthetic data generation from the Bayesian nonparametric model introduced in \Cref{pbd:sec:bnp}. In particular, we investigate (i) the role of the hyperparameters $\mass,\conc,\discount$  as well as (ii) the effect of sequencing depth on the rate at which rare variants appear in draws from the model.

\begin{figure}
    \centering
    \includegraphics[width=\textwidth,height=\textheight,keepaspectratio]{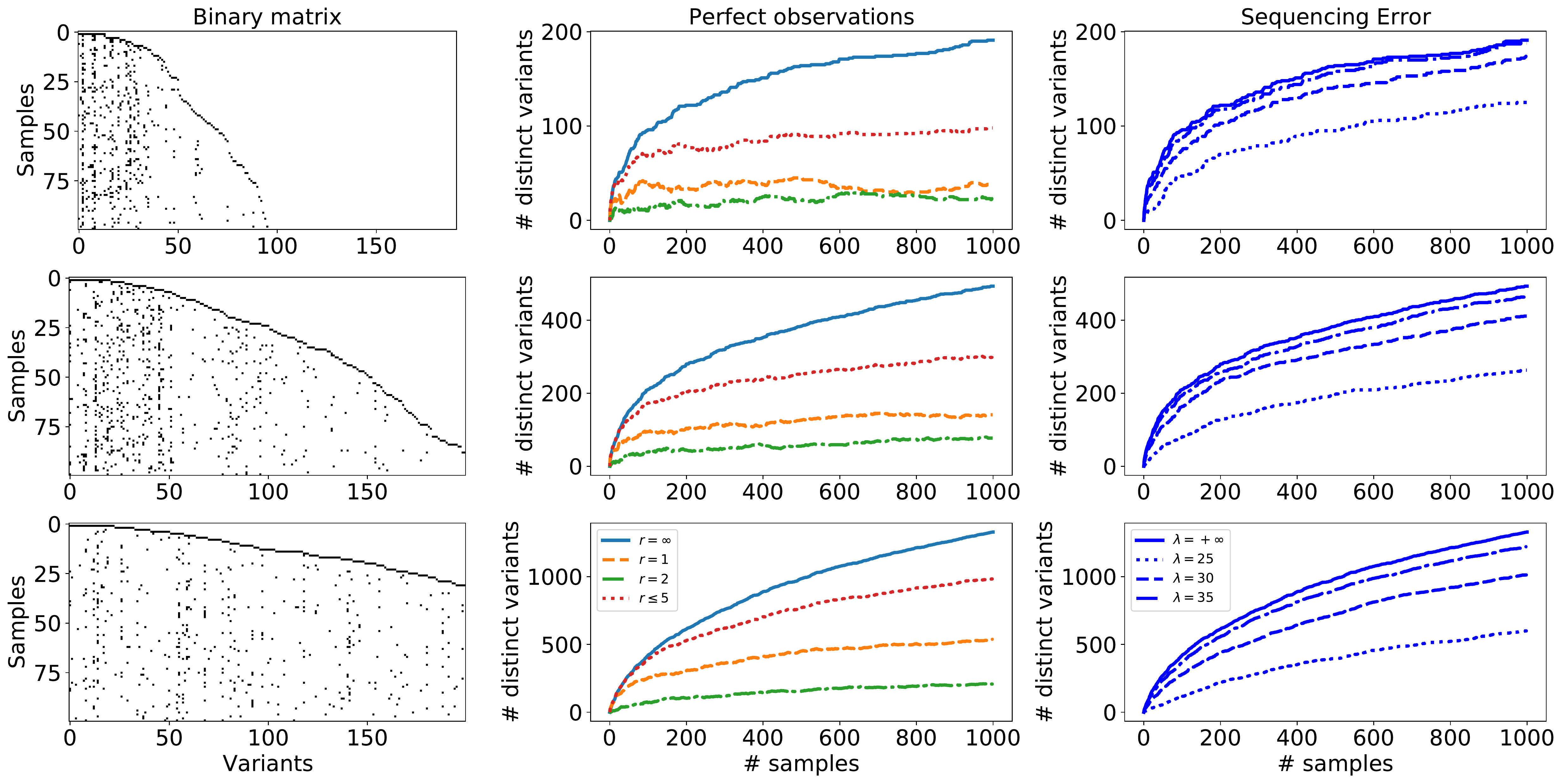}
    \caption{Predictive behavior of the $\tBP$-Bernoulli model (\Cref{pbd:eq:model_bnp}) under different choices of the hyperparameters. Each row refers to different choices of the model's hyperparameters. For each row, the subplot in the left column shows the binary matrix $X = [X_{n,\ell}]$ of Bernoulli processes, in which $X_{n,\ell} = 1$ (black square) if the $n$-th observation shows variation at locus $\ell$. The subplot in the central column shows instead the number of distinct variants present in the underlying binary matrix (vertical axis) as a function of the sample size (horizontal axis). Different colors and different line styles are used to denote different frequencies (blue: all variants, orange: $r =1$, green: $r=2$, red: $r \le 5$) at which variants occur. Last, the subplot in the right column reports the total number of observed distinct variants, under sequencing error, as the sequencing depth $\lambda$ varies. The underlying hyperparameters are $\bm{\xi} = [5,5,0.1]$ (first row), $\bm{\xi} = [8,7,0.2]$ (second row) and $\bm{\xi} = [10,15,0.3]$ (third row). }
    \label{pbd:fig:ibp_generative_viz}
\end{figure}

First, we show in \Cref{pbd:fig:ibp_generative_viz} simple draws from the three-parameter beta Bernoulli model, for different configurations of the model's hyperparameters and sequencing depth. In order to draw Bernoulli processes $X_n$, $n=1,2,\ldots$, we here resort to the exact ``marginal scheme'' (the Indian buffet process [IBP] \citep{thibaux2007hierarchical,teh2009indian}). A draw from the model is a sparse binary matrix, with a random number of columns (variants). The rate at which the number of columns grows as we increase the number of rows (samples) is governed by the hyperparameters of the process (see, e.g., \citet[Proposition 2]{masoero2021more}). When we also consider the effect of sequencing errors, lower sequencing depths  prevent us from uncovering all the underlying genetic variation due to sampling noise. In particular, we might miss a large fraction of \emph{rare} variants. 

\begin{figure}
    \centering
    \includegraphics[width=\textwidth,height=\textheight,keepaspectratio]{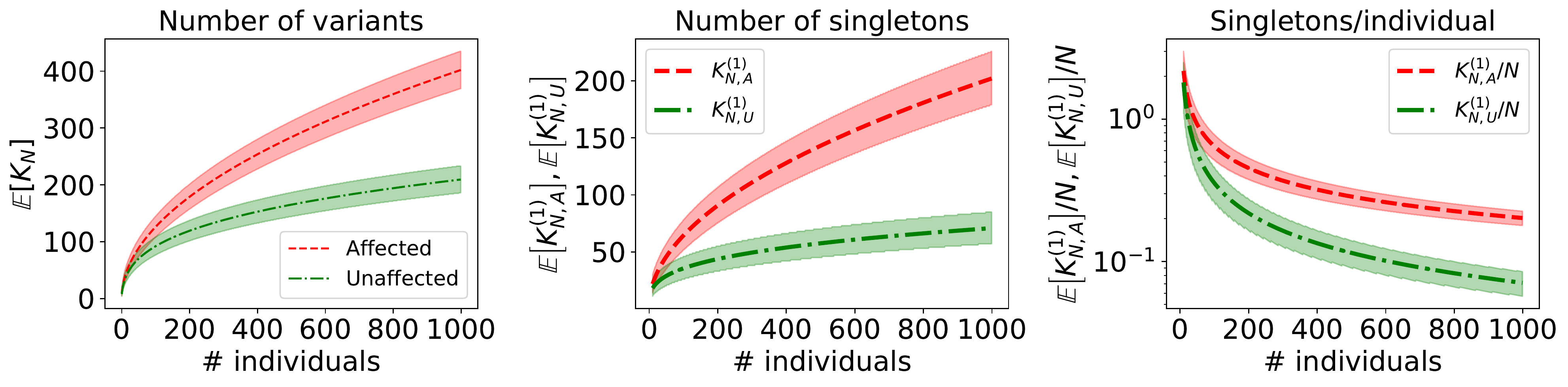}
    \caption{Predictive behavior of the $\tBP$-Bernoulli model (\Cref{pbd:eq:model_bnp}) in the absence of sequencing error. Here, $\bm{\xi}_A = [10, 0.5, 0.1]$ and  $\bm{\xi}_U = [8, 0.3, 1]$. In the left subplot, we show for both configuration of parameters the expected behavior of the number of distinct variants (vertical axis), $K_N$, as $N$ increases (horizontal axis). In the central subplot, we show  the expected number of singletons (vertical axis) as $N$ increases (horizontal axis). Last, in the right subplot, we show the expected number of singletons \emph{per individual} (vertical axis) as $N$ increases (horizontal axis). Shaded bands report $95\%$ credible intervals.}
    \label{pbd:fig:ibp_generative_0}
\end{figure}

Next, we move on to considering two distinct populations, ``affected'' $A$ and ``unaffected'' $U$, each population with its own set of parameters. For these experiments, we fix $\bm{\xi}_A = [10, 0.5, 0.1]$ and $\bm{\xi}_U = [8, 0.3, 1]$. While these choices are arbitrary,  they are such that the null hypothesis $H_0$ is enforced. Namely, draws from the first set of parameters $\bm{\xi}_A$ give rise (in expectation) to a larger number of rare variants than the corresponding draws given $\bm{\xi}_U$, as depicted in \Cref{pbd:fig:ibp_generative_0}.

\begin{figure}
    \centering
    \includegraphics[width=1\textwidth,keepaspectratio]{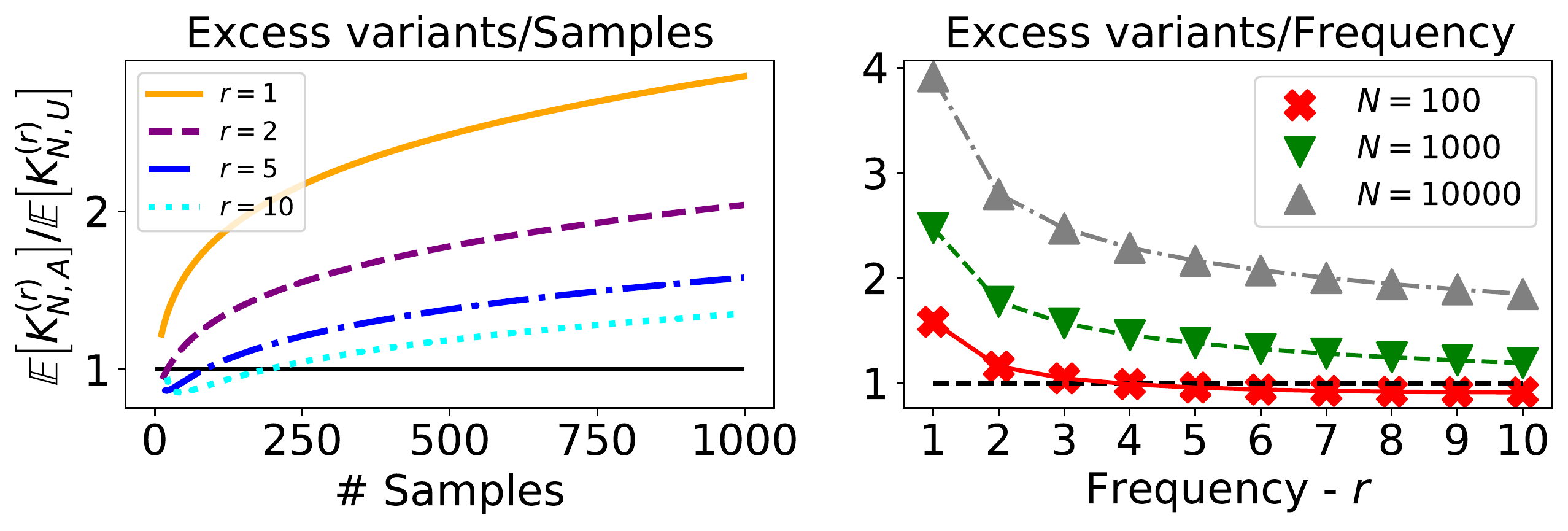}
    \caption{Excess variants ratios for the $\tBP$-Bernoulli model (\Cref{pbd:eq:model_bnp}). Here, like in \Cref{pbd:fig:ibp_generative_0},  $\bm{\xi}_A = [10, 0.5, 0.1]$ and  $\bm{\xi}_U = [8, 0.3, 1]$. In the left subplot, we show how the excess variants ratio $\epsilon_0^{(N,r)}$ (vertical axis)  changes as a function of the sample size $N$ (horizontal axis). Different lines show results for different frequencies. In the right subplot, for a fixed sample size $N$, we show on the vertical axis the excess variants ratio between the affected and unaffected population as we vary the variants' frequency (horizontal axis). Different lines correspond to different sample sizes $N\in\{10^2, 10^3, 10^4\}$. In both subplots, we place a horizontal line at $1$. This corresponds to the case in which both subpopulations show the same number of variants.}
    \label{pbd:fig:ibp_generative_1}
\end{figure}

An important feature of our model, is that it allows us to control the rate of additional variants we expect to see in the affected population with respect to the unaffected population. Conditionally on fixed $N~\ge~ 0$, observations from the model, we denote by $U_N^{(M)}(\bm{\xi}_j)$ the number of new variants to be observed in $M$ new samples given $N$ samples in population $j \in \{A,U\}$ --- where we emphasize the dependence of this quantity on the hyperparameter $\bm{\xi}_j$. Similarly, we let $U_N^{(M,r)}(\bm{\xi}_j)$ denote the number of new rare variants, observed exactly $r$ times in $M$ new samples given the $N$ original ones. Then, we define the ``excess variants ratio'' for the total and rare number of variants respectively to be
\[ 
    \epsilon_N^{(M)}:=\frac{U_N^{(M)}(\bm{\xi}_A)}{U_N^{(M)}(\bm{\xi}_U)}, \quad \text{and} \quad \epsilon_N^{(M, r)}:=\frac{U_N^{(M,r)}(\bm{\xi}_A)}{U_N^{(M,r)}(\bm{\xi}_U)}.
\]
Here $\epsilon_N^{(M)}$ is the total excess ratio, while $\epsilon_N^{(M, r)}$ is the excess ratio of variants appearing with frequency $r$. We show in \Cref{pbd:fig:ibp_generative_1} how different parameter specifications lead to different excess variants ratios (see \Cref{pbd:fig:ibp_generative_1}).

\begin{figure}
    \centering
    \includegraphics[width=1\textwidth,keepaspectratio]{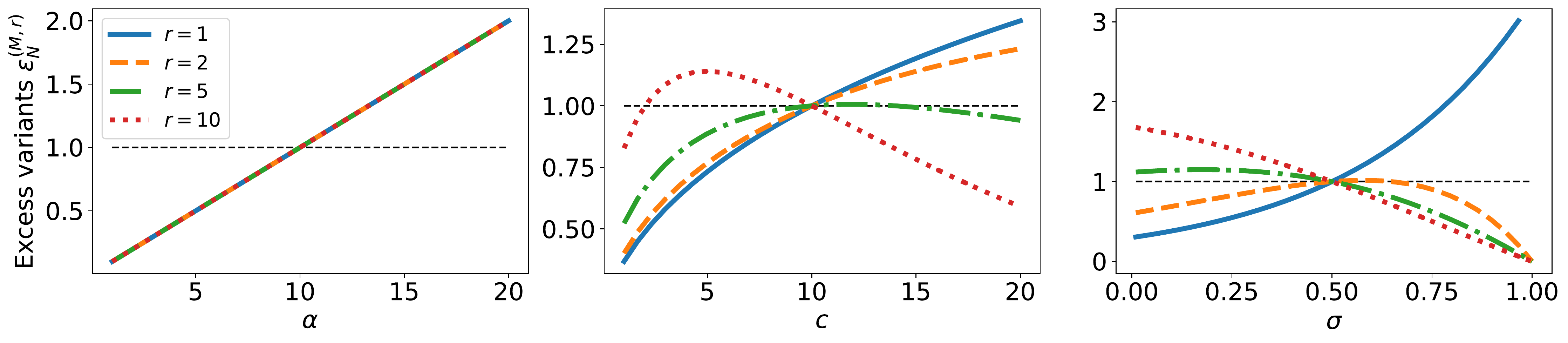}
    \caption{Hyperparameters' role in controlling excess variants ratios for the $\tBP$-Bernoulli model (\Cref{pbd:eq:model_bnp}). Here, we consider fixed $N = 0, M = 100$. In all three subplots, we fix the hyperparameters of the unaffected subpopulation to be $\bm{\xi}_U = [10, 10, 0.5]$. In turn, in each subplot, we let two of the three hyperparameters governing the affected subpopulation coincide with $\bm{\xi}_U$, and consider how the excess ratio of rare variants varies as we vary the third hyperparameter (left: $\mass \in [1,20]$, center: $\conc \in [1,20]$, right: $\discount \in [0,1)$). We report on the vertical axis the excess variants ratio $\epsilon_0^{100,r}$, for different choices of $r \in \{1,2,5,10\}$, as we vary the value of a single hyperparameter.}
    \label{pbd:fig:ibp_generative_2}
\end{figure}

We conclude our preliminary exploration by analyzing how each hyperparameter in the model affects the excess variants ratio (see \Cref{pbd:fig:ibp_generative_2}). We do so by comparing the value of $\epsilon_N^{(M)}, \epsilon_N^{(M,r)}$ as we vary in turn each of the three hyperparameters of $\bm{\xi}_A$, while keeping the other two parameters fixed, and equal to the value of the parameters in $\bm{\xi}_U = [10,10,0.5]$.
First, the ``mass'' parameter $\mass > 0$ simply scales the total variants' distribution, uniformly across the spectrum. We show in the left subplot of \Cref{pbd:fig:ibp_generative_2} how increasing $\mass$ has the effect of linearly increasing the excess variant ratio $\epsilon_N^{(M,r)}$, in the same way for all $r \ge 1$. The effect of the concentration parameter $\conc$, and of the discount parameter $\sigma$, are more subtle. In general, larger values of $\conc,\discount$ tend to favor a larger number of rare variants, but fewer common variants. That is, when $c, \sigma$ are large, the majority of variants are rare, observed in few individuals, while common variants are less present.

Overall, the preliminary analyses presented here display how the Bayesian nonparametric model introduced in \Cref{pbd:sec:bnp} is a flexible model, able to capture a wide range of data generating regimes.
\subsection{Power analysis with the Bayesian nonparametric model} \label{pbd:sec:exp_pow}
We now move to the power analysis for the rare variants burden test introduced in \Cref{pbd:sec:burden_tests}. We focus on prior analyses for the ``balanced samples'' case. I.e.\ we always consider $N_A = N_U = 0$ (no prior samples), and $M_A = M_U$. Further, we assume that the cost is linear in the samples and the depth --- that is, there is no library preparation cost, and the cost of sequencing $m$ samples at depth $\lambda$ is given by $c(m,\lambda) = m \lambda$ (i.e., $\kappa_0 = 0,\kappa_1=1$ in the RHS of \Cref{pbd:eq:power_maximization}).

\paragraph{Fixed design.} We start by considering the ``fixed design'' case (Q1). We here focus on singletons, i.e.\ the case $r=1$. We are interested in understanding how the statistical power of the burden test changes as a function of the number of samples collected, for a fixed sequencing depth. That is, we imagine that the sequencing depth of the experiment is fixed, and experimenters are interested in understanding how, for a given significance level, the power of their test will change as the sample size increases. Our simulations confirm the intuition that --- for the same number of samples --- a higher sequencing depth always allows to achieve a better power, when $H_0$ is true. 

\begin{figure}
    \centering
    \includegraphics[width=1\textwidth,keepaspectratio]{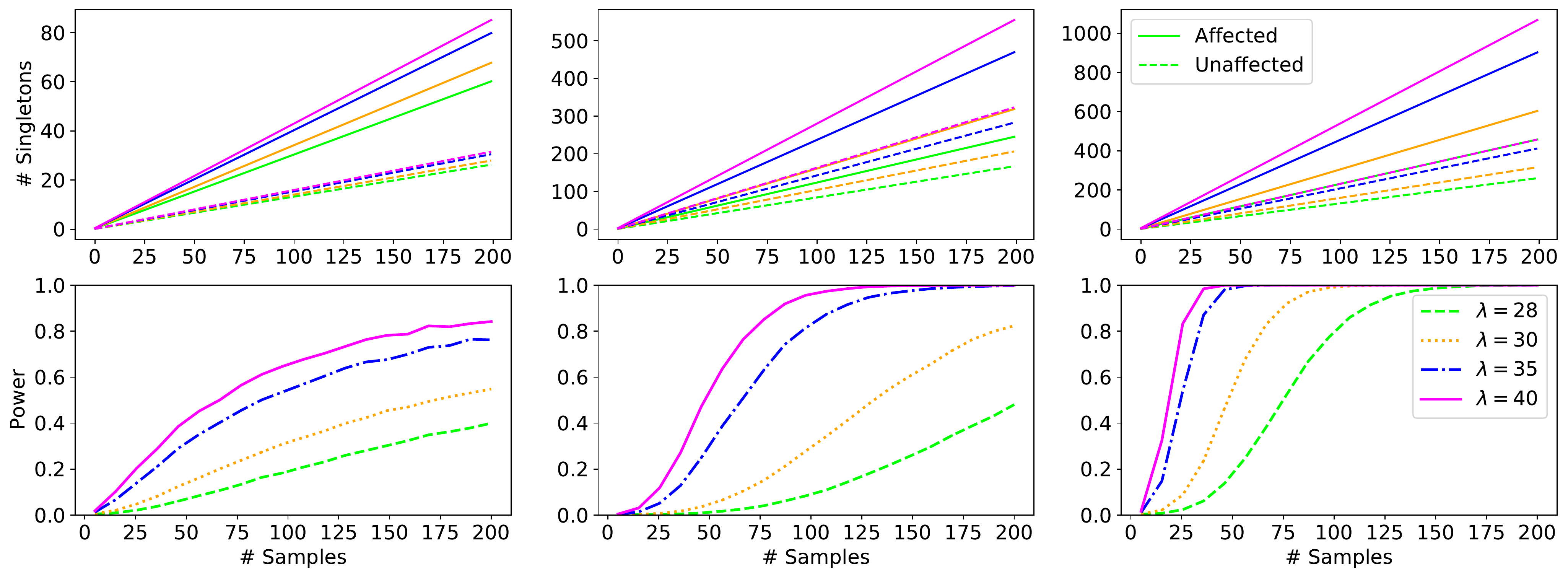}
    \caption{Power analysis under fixed design. In each column, we consider two different subpopulations (affected, and unaffected individuals). The data generating process of each group is driven by different sets of hyperparameters  (left: $\bm{\xi}_A = [10,4,0.2]$, $\bm{\xi}_U = [7,3,0.1]$, center: $\bm{\xi}_A=[15,8,0.5]$, $\bm{\xi}_U=[14,6,0.4]$, right: $\bm{\xi}_A = [24,12,0.5]$, $\bm{\xi}_U=[20,10,0.3]$).
    In each subplot, we show under fixed sequencing budget how different sequencing depth choices (different colored lines) affect the power of the singletons burden test as we increase the sample size (horizontal axis). In our experiments, we set the significance level (type-I error threshold) to be $10^{-4}$. }
    \label{pbd:fig:ibp_fix_design}
\end{figure}

The extent to which having a higher sequencing depth is important, however, depends on a number of factors. In particular, if the goal is to achieve a given level of power, we expect that different datasets will require different sequencing choices. For example, if the disease under study is characterized by a large excess ratio, relatively lower depth might suffice to achieve the desired power. Instead, if the excess ratio is close to one, higher depth might be needed to achieve the desired power. To highlight this,  we show in \Cref{pbd:fig:ibp_fix_design} results for three different couples of datasets of affected and unaffected subpopulations. Each couple of datasets was drawn from our model under different configurations of the hyperparameters. In each case, we enforce the null hypothesis $H_0$ to be true, by choosing the hyperparameters in such a way that the affected individuals show a larger number of singletons than the unaffected individuals. In general, and as expected, larger number of samples and higher sequencing depth always lead to improved power. However, the extent to which choosing a larger sequencing depth affects the power (hence, the usefulness) of the experiments, is greatly affected by the characteristics of the data: when (i) fewer variants are present and (ii) the excess variants ratio is small, a relatively higher depth is needed in order to achieve a given level of power.

\begin{figure}
    \centering
    \includegraphics[width=1\textwidth,keepaspectratio]{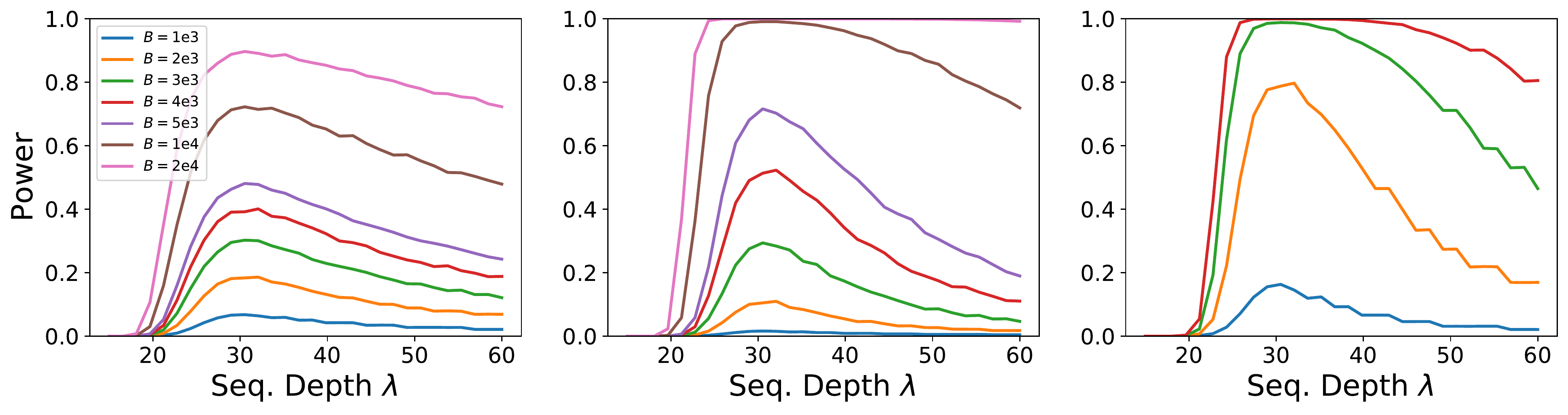}
    \caption{Power analysis under fixed budget. For each subplot, we consider two different subpopulations (affected, and unaffected individuals respectively) --- each group driven by different sets of hyperparameters (left: $\bm{\xi}_A = [10,4,0.2]$, $\bm{\xi}_U = [7,3,0.1]$, center: $\bm{\xi}_A=[15,8,0.5]$, $\bm{\xi}_U=[14,6,0.4]$, right: $\bm{\xi}_A = [24,12,0.5]$, $\bm{\xi}_U=[20,10,0.3]$).
    In each subplot, we report the power of the singletons burden test (vertical axis) as we increase the sequencing depth (horizontal axis), under a fixed budget constraint, the assumption of balanced samples (same number of samples from each subpopulation) and given the cost function $c(m,\lambda) = 2m\lambda$. Different colored lines report the power for a different budget as a function of the implicit feasible sample sizes $M_A=M_U$, as we vary $\lambda$. In our experiment, we set the significance level to be $10^{-4}$. }
    \label{pbd:fig:ibp_budget}
\end{figure}

\paragraph{Fixed budget.} We conclude this set of experiments with results for the other scenario of interest, the ``fixed budget'' case (Q2). Here, we consider instead the setting in which scientists have the ability to choose the sequencing depth, but are limited by a finite, fixed budget. As shown in \Cref{pbd:fig:ibp_budget}, a clear trade-off emerges. On the one hand, sequencing with a smaller depth allows scientists to collect larger cohorts of individuals, and potentially reveal large numbers of rare variants. However, if the sequencing depth is too low, many variants are missed due to noise in the sampling process. As a consequence, the associated power of the test is low. On the other hand, sequencing with a higher depth has the benefit that most variants, including rare ones, are detected in the sampling process. However, under a fixed budget, the cost of sequencing forces relatively smaller sample sizes. The best choice is then obtained by balancing between a sufficiently high sequencing depth, and large sampling cohort. Again, we point out that under different hyperparameters of the data generating process, different optimal choices of sequencing depth would emerge, even for the same budget. Our results are qualitatively in line with the findings of \citet[Figure 4 and 5]{rashkin2017optimal}. 

\section{Improvements of the Bayesian framework} \label{pbd:sec:bnp_hierar}
The model proposed in \Cref{pbd:sec:bnp} enjoys a number of benefits: its theoretical properties are well understood, it is amenable to straightforward computation, and it has been previously largely and successfully employed in the applied literature (see, e.g.\ \citet{griffiths2011indian} for a review). However, this model also suffers from a number of limitations for the task at hand:
\begin{enumerate}
    \item[(a)] Inability to jointly model a variant's occurrence in different subpopulations. In turn, this model is not useful for the hypothesis test $H_0^\star$, which is designed to test for variants that are specific to one subpopulation.
    \item[(b)] Inability to model diploid sequences. Specifically, in \Cref{pbd:sec:bnp}, we treat every observation as a \emph{binary} vector of variants, in which we simply record the presence or absence of a discrepancy with respect to the reference genome. In reality, the human genome is diploid. That is, each variant can appear in either one, two or no copies. As such, it would be better model by values in $\{0,1,2\}$.
\end{enumerate}
To overcome both these limitations, we now introduce an extension of the model proposed in \Cref{pbd:sec_app:bnp_uni}, building off the framework first proposed in \citet{masoero2018posterior}.

\subsection{A Bayesian nonparametric hierarchical model formulation}
We start by addressing the first limitation --- (a). This limitation is related to the choice of the prior distribution employed in \Cref{pbd:sec:bnp}: that is, the way in which we model the distribution of variants' frequencies. There, we proposed independent priors for the affected and unaffected subpopulations. In turn, this assumption does not allow us to reason about variants' frequencies jointly across multiple subpopulations. In particular, the model of \Cref{pbd:sec:bnp} does not allow us to test us for the null hypothesis $H_0^\star$.

To overcome this issue, we now formulate a richer hierarchical model. We assume that there exists a shared, latent distribution over the variants' frequencies. Again, just like we did in \Cref{pbd:sec:bnp}, we assume that the distribution over these frequencies follows a Poisson point process, whose rate measure is given by \Cref{pbd:eq:rate_meas}. We call the shared latent frequency of the $k$-th variant $\theta_{0,k}$. We collect the variants' frequencies, each with a corresponding label $\aloc_k$, in a random measure:
\[
    \Theta_0 = \sum_k\theta_{0,k}\delta_{\aloc_k} \sim \tBP(\alpha_0, c_0, \sigma_0).
\]
Let now $J \ge 1$ be the number of subpopulations of interest. In what follows, we typically consider the case $J=2$, for the two subpopulations of affected and unaffected individuals, as in our previous example. Then, for each subpopulation $j = 1,\ldots,J$, we assume that for every variant $\aloc_{k}$ in the shared prior measure $\Theta_0$, there exists a population-dependent probability of observing a variant at the $k$-th locus in the $j$-th subpopulation. We call this probability $\theta_{j,k}$, and assume that it is characterized by the distribution already given in \Cref{pbd:eq:h_freq}:
\begin{align*} 
    \theta_{j,k} \mid \Theta_0 \sim \Betadist\left\{a_j \theta_{0,k}, b_j(1-\theta_{0,k})\right\}.
\end{align*}
That is, for every variant, the probability of observing it in population $j$ (i) can vary in different populations but (ii) depends on the underlying shared parameter $\theta_{0,k}$.
For each population, we collect these probabilities, together with the labels $\aloc_k$ in another random measure, $\Theta_j$:
\begin{align*}
    \Theta_j \mid \Theta_0 = \sum_k \theta_{j,k} \delta_{\aloc_k},
\end{align*}
and we denote the distribution as the vector of random measures $\Theta_1,\ldots,\Theta_J$ as
\begin{align}
    [\Theta_1,\ldots,\Theta_J] \mid \Theta_0 \sim \hBP(a_{1:J}, b_{1:J}; \Theta_0). \label{pbd:eq:bnp_multivariate}
\end{align}
We refer to the vector of random measures $[\Theta_1,\ldots,\Theta_J]$, as a ``hierarchical three-parameter Beta process'' [h3BP].

\subsection{Diploid observations: beyond Bernoulli processes}

Next, we move on to address the second limitation --- (b). This limitation is instead due to the choice of the likelihood model employed in \Cref{pbd:sec:bnp}. Namely, the choice of a simple Bernoulli process for the observational data. While the assumption of binary observations in the context of genomic discovery has been widely used in the literature \citep{ionita2009estimating, ionita2010optimal, gravel2014predicting, zou2016quantifying,chakraborty2019using}, in reality the human genome is diploid. I.e.\ it would better modeled by a variable that can take on three different values: either no variation at all (``aa'', homozygous reference), variation at one of the two copies (``aA'' or ``Aa'', heterozygous), or variation on both copies (``AA'', homozygous variant). Then, for each site $\ell$ and each observation $n$ in population $j$, the underlying genotype expression $g$ would better be represented by a ``ternary'' variable, $g \in \{0,1,2\}$, where $g=0$ when the site is non-variant homozygous (``aa''), $g= 1$ for a heterozygous site (``aA'' or ``Aa'') and $g = 2$ for a variant homozygous site (``AA''). 

Following standard assumptions in the genomics literature \citep{wigginton2005note,mayo2008century}, we assume Hardy-Weinberg equilibrium [HWE] proportions. That is, conditionally on the variant's frequency (in population $j$) $\theta_{j,k}$, in the absence of sequencing errors,
\begin{align}
    \PP(X_{j,n,k}=g \mid \theta_{j,k}) = 
        \begin{cases} 
            (1-\theta_{j,k})^2 &\mbox{ if } g = 0 \\
            2(1-\theta_{j,k})\theta_{j,k} &\mbox{ if } g = 1 \\
            \theta_{j,k}^2 &\mbox{ if } g = 2  
        \end{cases}.\label{pbd:eq:hwe}
\end{align}
The $n$-th observation in the $j$-th subpopulation is then a generalization of the Bernoulli process --- a multinomial process, denoted as
\[
    X_{j,n} \mid \Theta_j = \sum_{\ell \ge 1} X_{j,n,\ell} \delta_{\aloc_\ell} \sim \MultiP(\Theta_j, h(\cdot)).
\]
Here $h:[0,1]\to [0,1]^3$ is the function which takes as input a variant's probability $\theta \in [0,1]$ and returns the corresponding Hardy-Weinberg proportions,
\[
    h(\theta) = \begin{bmatrix} (1-\theta)^2, & 2\theta(1-\theta), & \theta^2 \end{bmatrix}^\top,
\]
and each variant appears according to the HWE proportions given in \Cref{pbd:eq:hwe}, i.e.\
\[
    X_{j,n,\ell} \mid \Theta_j \sim \Multi\left((1-\theta_{j,\ell})^2, 2\theta_{j,\ell}(1-\theta_{j,\ell}), \theta_{j,\ell}^2\right).
\]

\subsection{Sequencing error in the hierarchical diploid model} \label{pbd:sec:h_seq_error}
Next, we have to discuss how the presence of noise in sequencing affects the results from our model. We here assume the same error model as the one introduced in \Cref{pbd:sec:model_noise}, with the only difference that now, whenever sufficiently many reads are produced, the correct haploid or diploid variant  is called. I.e., again 
\[
    Z_{j,n,\ell} = \ind(C_{j,n,\ell,\text{noerror}}>D)X_{j,n,\ell}.
\]
Hence, it follows
\[
    X_{n,j} \mid \Theta_j \sim \MultiP(\phi\Theta_j, h(\cdot)),
\]
i.e.
\[
    \PP(X_{n,j} = g \mid \Theta_j, \lambda, D) = 
        \begin{cases} 
            (1-\phi\theta_{j,k})^2 &\mbox{ if } g = 0 \\
            2(1-\phi\theta_{j,k})\phi\theta_{j,k} &\mbox{ if } g = 1 \\
            (\phi\theta_{j,k})^2 &\mbox{ if } g = 2  
        \end{cases},
\]
with $\phi = \phi(\lambda, D, p_{err}, T) = 1-\PP(Y>D)$, for $Y\sim \Pois((1-p_{err})\lambda)$. To sum up,
\begin{align}
\begin{split}
    \Theta_0 &\sim \tBP(\mass_0, \conc_0, \discount_0), \\
    [\Theta_1,\ldots,\Theta_J] \mid \Theta_0 &\sim \hBP(a_{1:J}, b_{1:J} ; \Theta_0),\\ X_{j,n} \mid \Theta_j &\sim \MultiP(\phi \Theta_j, h(\cdot)).
\end{split} \label{pbd:eq:h_model}
\end{align}

We discuss possible extension to alternative variant calling rules, e.g.\ based on genotype likelihoods, in \Cref{pbd:sec:discussion}.

\subsection{Data generation and power trade-offs in the hierarchical model}

We now have all the necessary ingredients to discuss how this hierarchical model can prove useful for power trade-off considerations. First, recall the
non-hierarchical setting. The experiments for power maximization presented in \Cref{pbd:sec:experiments} exploit the analytic predictive distribution for the number of new, yet unobserved (potentially, rare) variants, given in \Cref{pbd:prop:k_tons_error}. In principle, we could do the same in the hierarchical setting: i.e.\ we could exploit the posterior characterization provided in \citet[Theorem 2]{masoero2018posterior} and derive analogous analytic formulae for the posterior predictive distribution of the number of new, rare variants in the hierarchical setting. In practice, however, in the hierarchical setting such formulae do not enjoy the same practicality and simplicity of their univariate counterpart. In particular, the hierarchical model here adopted is such that the number of new, rare variants to be observed in either number of subpopulations is still Poisson distributed, but the parameter of such distribution is only available in an integral form, due to the lack of conjugacy in the hierarchical model. We find that computation of such integral is prone to numerical error. 

Therefore, in our experiments, instead of relying on the analytic characterization of the number of new rare variants, we resort to simulation: specifically, we approximate the expected number of new rare variants by Monte Carlo approximation: because the hierarchical model in \Cref{pbd:eq:h_model} provides us with a full generative scheme, we can repeatedly draw from the model, and approximate the number of new rare variants by Monte Carlo averaging.
\section{Experiments from the hierarchical model} \label{pbd:sec:experiments_hierar_2}

We now move on to the empirical evaluation of the properties of the Bayesian hierarchical model introduced in \Cref{pbd:sec:bnp_hierar}. First, in \Cref{pbd:sec:h_exp_viz}, we start by analyzing the effect of the additional parameters $a_j, b_j$, to understand how the hierarchical model differs from the three-parameter beta-Bernoulli used in \Cref{pbd:sec:experiments}. In what follows, we adopt the same default specifications as in \Cref{pbd:sec:experiments} for the threshold parameter $D=30$ and the error $p_{error} = 0.05$.

\subsection{New, rare and excess variants in the Bayesian nonparametric hierarchical model} \label{pbd:sec:h_exp_viz}

\begin{figure}
    \centering
    \includegraphics[width=\textwidth,height=\textheight,keepaspectratio]{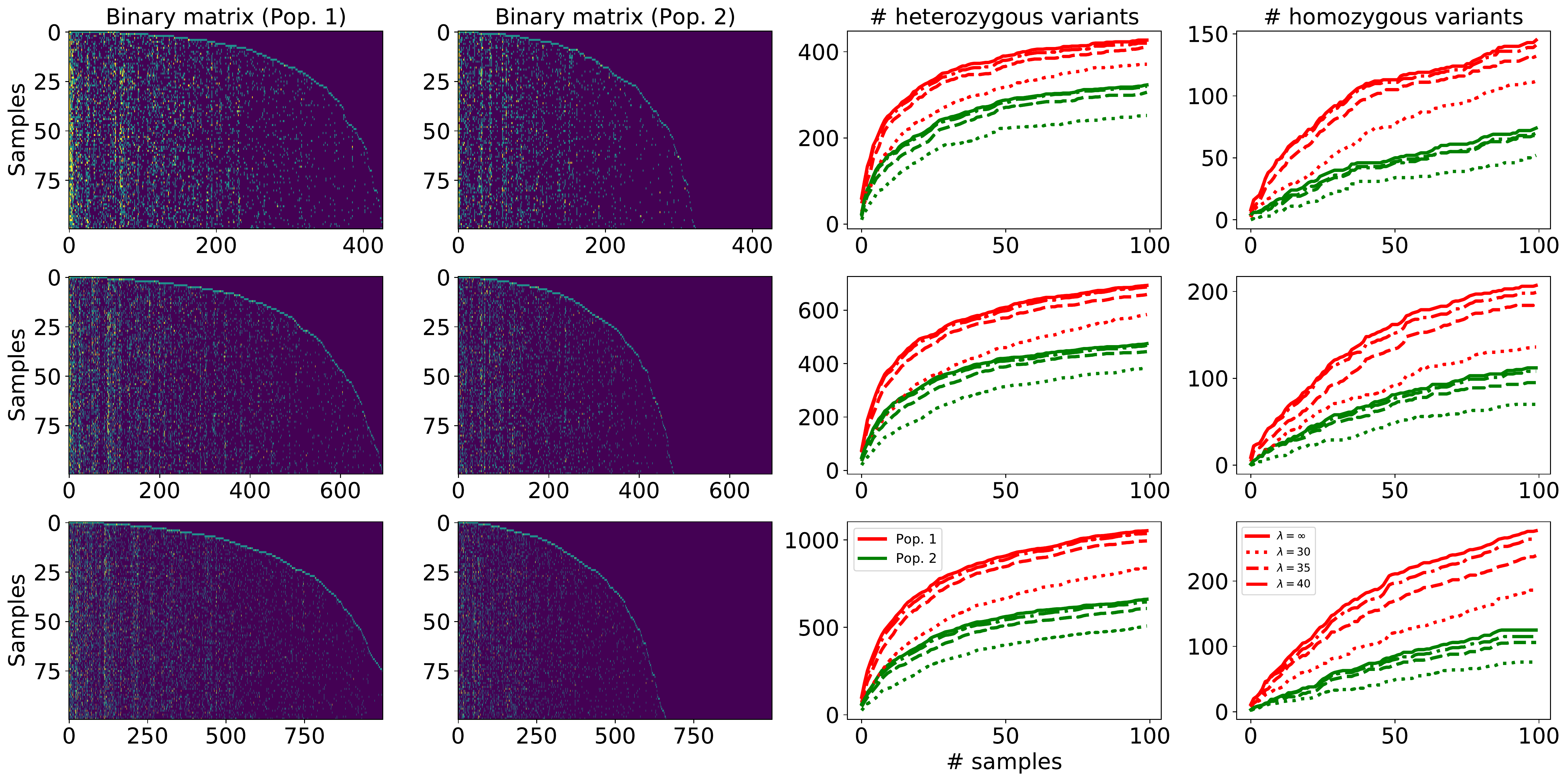}
    \caption{Predictive behavior for the $\hBP$-Multinomial model under different choices of the generative parameters. Here, we focus on the case of two populations. The two left columns show the genotype matrices $X^{(j)} = [X_{j,n,k}] \in \{0,1,2\}^{N_j\times K}$, in which $X_{j,n,k} = 0$ (purple) if the $n$-th observation shows no variation at locus $k$, $X_{j,n,k} = 1$ (green) if there is a heterozygous variant, and  $X_{j,n,k} = 2$ (yellow) if the variant is homozygous. The third and fourth columns show the total number of distinct homozygous and heterozygous variants present in the underlying genotype matrices $X_{j,n,k}$. We show in the third column the number of loci in which at least one heterozygous variant is detected (vertical axis) as a function of the sample size (horizontal axis), while in the fourth column we display the same quantity, now for heterozygous variants. Different colors are used to refer to different populations. Different line-styles are used to denote different sequencing depths. Different rows refer to different choices of the underlying hyperparameters $\bm{\xi}_0 = [\mass_0, \conc_0,\discount_0]$ --- first row: $\bm{\xi}_0 = [20,10,0.1]$ (first row), $\bm{\xi}_0 = [25,15,0.2]$ (second row) and $\bm{\xi}_0 = [30,20,0.3]$ (third row). For all rows, we set $a_1 = 200, a_2 = 100$, and $b_1 = b_2 = 100$.}
    \label{pbd:fig:hibp_generative_viz}
\end{figure}

The hierarchical model introduced in \Cref{pbd:sec:bnp_hierar} allows us to jointly model occurrence of variants in multiple subpopulations. From a modeling standpoint, this is desirable, as the same variant could share the same role, even in different subpopulations. In our discussion, we let subpopulation 1 be the ``cases'' (affected subpopulation), and subpopulation 2 be the controls (unaffected subpopulation). To understand the properties of the model, we start by analyzing the effect of the hyperparameters $\mass_0, \conc_0, \discount_0$ on the data generating process. These parameters directly govern the underlying base measure $\mfreq_0$. Similarly to what we discussed in \Cref{pbd:sec:experiments}, also in the hierarchical setting these parameters control the rate of growth of the total number of variants, similarly as the univariate counterpart. Larger values of $\mass_0$ scale linearly the number of variants we expect to see, while larger values of $\conc_0, \discount_0$ tend to favor a larger number of rare variants. To provide intuition, we show simple draws from the hierarchical model in  \Cref{pbd:fig:hibp_generative_viz}. 

Next, we move on to the discussion of the hyperparameters $a_j, b_j$, for $j=1,\ldots,J$. These parameters have an important role for determining the distribution of the variant's distribution in each subpopulation: for a given locus $\ell$ with underlying common base frequency $\theta_{0,\ell} \in [0,1]$, from the representation in \Cref{pbd:eq:h_freq}, it follows that
\begin{align}
    \E[\theta_{j,k} \mid \theta_{0,k}] = \frac{a_j\theta_{0,k}}{b_j + (a_j-b_j)\theta_{0,k}}, 
\end{align}
and
\begin{align}
    \mathrm{Var}[\theta_{j,k} \mid \theta_{0,k}] = \frac{  {a_jb_j\theta_{0,k}(1-\theta_{0,k})}  }{\left(b_j + (a_j-b_j)\theta_{0,k}\right)^2\left(1+\theta_{0,k} + (a_j-b_j)\theta_{0,k}\right)}.
\end{align}

\begin{figure}
    \centering
    \includegraphics[width=\textwidth,height=\textheight,keepaspectratio]{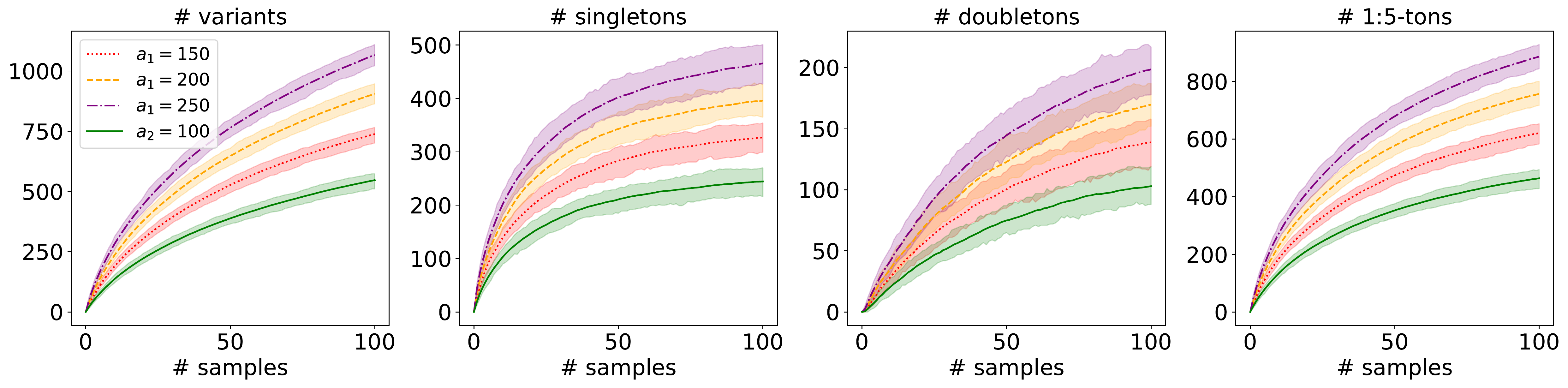}
    \caption{Predictive behavior for the $\hBP$-Multinomial model under different choices of the generative specifications. Here, we fix hyperparameters for the underlying shared frequencies $\mfreq_0$: $\alpha_0=10, c_0=4,\sigma_0 =0.7$. We fix hyperparameters $b_1=a_2=b_2=100$, and consider three possible values for $a_1 \in \{150,200,250\}$. For each configuration of the hyperparameters, we report the predictive behavior by generating from the model.}
    \label{pbd:fig:hibp_generative_viz_2}
\end{figure}

\begin{figure}
    \centering
    \includegraphics[width=\textwidth,height=\textheight,keepaspectratio]{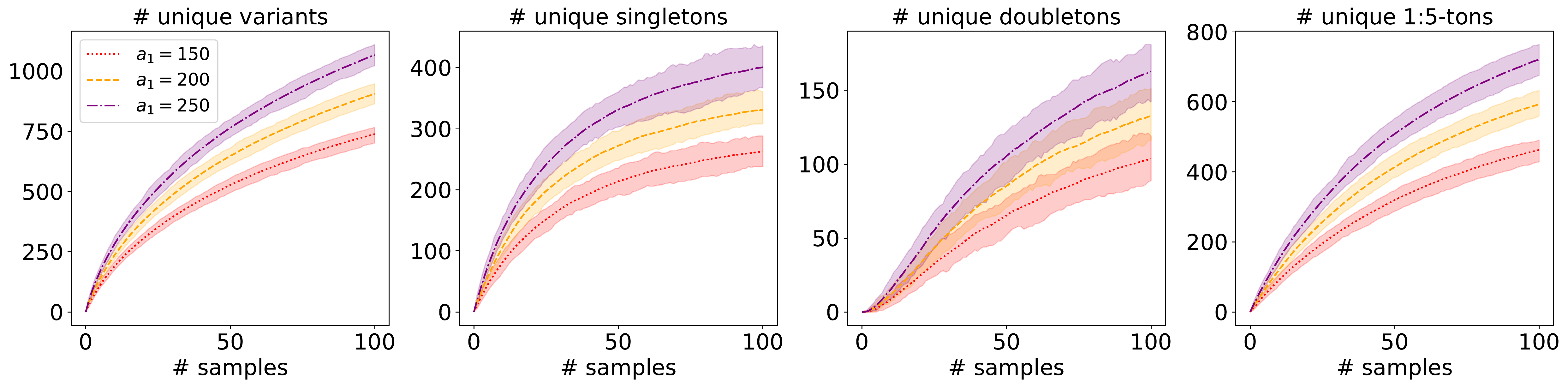}
    \caption{Predictive behavior for the $\hBP$-Multinomial model under different choices of the generative parameters. Under the same model specifications as in \Cref{pbd:fig:hibp_generative_viz_2}, we now focus on ``exclusive'' variants --- namely variants that appear only in population $j=1$, and do not appear in population $j=2$.}
    \label{pbd:fig:hibp_generative_viz_3}
\end{figure}

That is to say, the parameters $a_j,b_j$ control the dispersion of the hierarchical frequencies with respect to the underlying ``base'' frequencies. We show in \Cref{pbd:fig:hibp_generative_viz_2,pbd:fig:hibp_generative_viz_3} how different choices of $a_j$ can impact the rate of growth of distinct variants, as well as rare variants (singletons, doubletons). In \Cref{pbd:fig:hibp_generative_viz_2} we focus on variants that appear in the affected population, and also, potentially, in the unaffected subpopulation. Instead, we focus on those variants that appear \emph{exclusively} in the affected subpopulation in \Cref{pbd:fig:hibp_generative_viz_3}. Different choices of the hyperparameters have a great impact on the rate of growth of the number of variants; this shows that our model has the ability to capture a wide range of different data generating behaviors.

\subsection{Power analysis with the hierarchical model} \label{pbd:sec:h_power}

We now move to the power analysis for the hierarchical model. Again, we focus on the two questions --- Q1 and Q2 --- already investigated with the univariate model.

In \Cref{pbd:fig:hibp_fixed}, we consider the ``fixed-design'' case (Q1), in which we analyze how the power of the singleton burden test --- now for $H_0^\star$ (\Cref{pbd:eq:simple_burden_test_unique}) --- changes as we simultaneously increase the sizes of the control and the affected subpopulations, under a fixed sequencing depth. Larger sample sizes, and higher sequencing depths, are never harmful for power. However, the rate at which the power increases depends on the underlying hyperparameters of the process. When working with real data, therefore, different sequencing strategies could prove more or less effective in different contexts. In particular, if the experiment under study has the goal of achieving a desired level of power, our framework could allow practitioners to provide estimates of the budget needed in order to achieve the desired power level.

\begin{figure}
    \centering
    \includegraphics[width=\textwidth,height=\textheight,keepaspectratio]{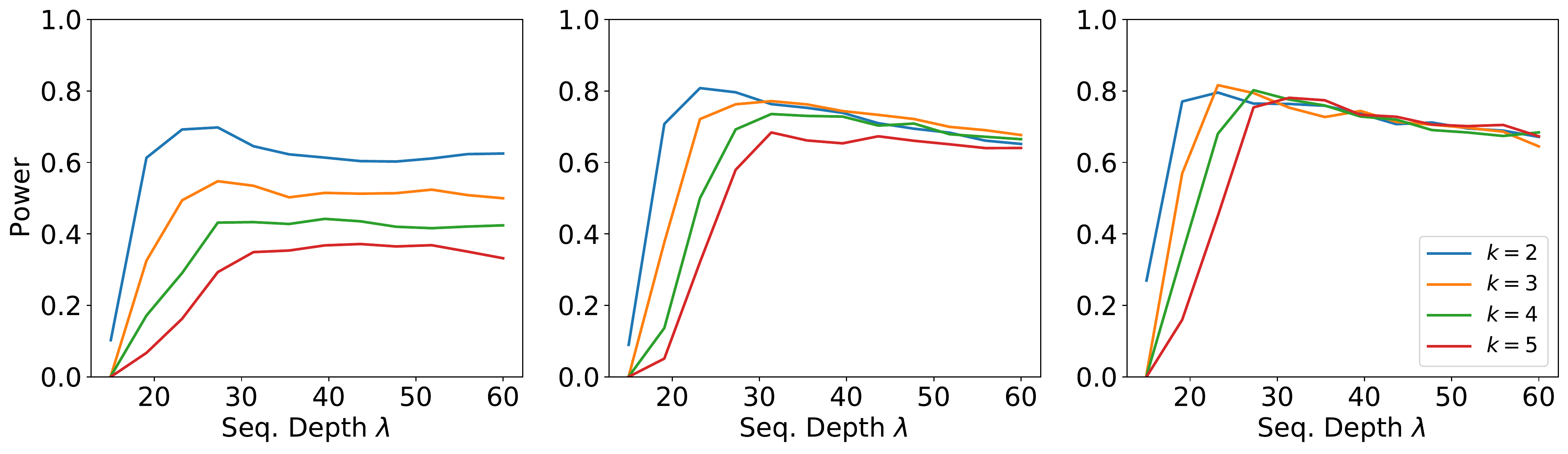}
    \caption{Under the same generative specifications as \Cref{pbd:fig:hibp_budget}, we report results for the $k$-tons burden test, for $k = 2$ (blue), $k = 3$ (orange), $k = 4$ (green), and $k = 5$ (red) under the same, fixed budget of $B=5000$ units, where the cost $c(m,\lambda) = m\lambda$. }
    \label{pbd:fig:hibp_budget_ktons}
\end{figure}

Next, we move on to the analysis of the second problem considered in our introduction --- the power maximization under a fixed budget (Q2). We here consider both singletons, as in \Cref{pbd:sec:experiments}, as well as $k$-tons --- i.e.\ all variants appearing at most $k$ times --- for $k \in \{2,3,4,5\}$. Under our assumptions, we find that the optimal choice of the sequencing depth --- i.e.\ the depth maximizing the corresponding test power under a fixed budget constraint:
\begin{itemize}
    \item is relatively insensitive to the available budget: that is, under a given set of experimental hyperparameters for the data generating process, the optimal sequencing depth will tend to be similar for different available budgets. See e.g.\ \Cref{pbd:fig:hibp_budget}. These findings are in line with \citet[Figures 4,5]{rashkin2017optimal}.
    \item depends more heavily on the type of the test. We find that (i) tests for extremely rare variants generally achieve higher power than tests for relatively less rare variants, and that (ii) the optimal sequencing depth is typically lower for tests of extremely rare variants (see \Cref{pbd:fig:hibp_budget_ktons}).
\end{itemize}
The intuition behind this behavior has to be sought in the properties of the model's frequencies distribution: the underlying three-parameter beta process prior suggests that most variants will be extremely rare. Therefore, when testing for extremely rare variants, the test has sufficient power even when the sequencing depth is relatively lower, just because on average the affected population will reveal a larger number of singletons with respect to the unaffected population. On the contrary, when testing for relatively more frequent variants (e.g., $k$-tons), a larger depth is needed in order to capture a significant discrepancy between the affected, and unaffected population.

\end{document}